\documentclass[a4paper,11pt]{article}

\usepackage{jinstpub} 

\title{Systematic uncertainties in integrated luminosity measurement at CEPC}

\author[a,1]{I. Smiljanic,\note{Corresponding author.}}
\author[a]{I. Bozovic Jelisavcic,}
\author[a]{G. Kacarevic,}
\author[a]{N. Vukasinovic,}
\author[a]{I. Vidakovic,}
\author[a]{and V. Rekovic}

\affiliation[a]{Vinca Institute of Nuclear Sciences - National Institute of the Republic of Serbia, University of Belgrade,\\M. Petrovica Alasa 12-14, Belgrade, Serbia}

\emailAdd{i.smiljanic@vin.bg.ac.rs}

\abstract{The very forward region is one of the most challenging regions to instrument at a future $e^{+}e^{-}$ collider. At CEPC, machine-detector interface includes, among others, a calorimeter dedicated for precision measurement of the integrated luminosity at a per mill level or better. Here we review a feasibility of such precision, from the point of view of systematic effects arising from luminometer mechanical precision and positioning, beam-related requirements and physics background from two-photon processes. The impact of the beam energy spread and its uncertainty on the integrated luminosity precision is also discussed, as well as the achievable beam energy spread precision with the post-CDR CEPC beams.}


\arxivnumber{2010.15061} 

\begin{document}
\maketitle
\flushbottom

\section{Introduction}
\label{sec:intro}

To achieve precision required for realization of the CEPC physics program, relative uncertainty of the integrated luminosity measurement should be of order of $10^{-4}$ at the $Z^{0}$ pole (91.2 GeV) and of order of $10^{-3}$ at 240 GeV $e^{+}e^{-}$ beam center-of-mass (CM) energies. The method of integrated luminosity measurement at CEPC, as well as the machine parameters, detector concept, machine-detector interface (MDI) and physics performance, is described in \cite{CEPC_CDR}. Precise reconstruction of position and energy of electromagnetic showers generated by the Bhabha scattering at a high-energy $e^{+}e^{-}$ collider can be achieved with finely granulated luminometer \cite{FCAL}. However, there is a long list of systematic uncertainties in integrated luminosity measurement, that includes detector related uncertainties, beam related uncertainties and uncertainties originating from physics and from machine related interactions. Here we review the effects of detector and beam related uncertainties, namely uncertainties on the luminometer mechanical positioning and size and uncertainties on the beam energy, beam synchronization and interaction point (IP) position. Also, we review the uncertainty originated from miscount of two-photon background (Section \ref{sec:sec3}). Motivated by \cite{Janot}, in Section \ref{sec:sec4} we discuss the possibility of CEPC beam energy spread (BES) determination with the post-CDR beam parameters \cite{Gao} and its impact on the integrated luminosity precision. In addition, we discuss the impact of the estimated BES precision on several relevant electroweak observables at the $Z^{0}$ pole:  $Z^{0}$ production cross-section, mass and width. Uncertainties from beam-beam interactions and beam-gas scattering are not discussed in this paper. 

\section{Forward region at CEPC}
\label{sec:sec2}
\begin{figure}[htbp]
\centering 
\includegraphics[width=.60\textwidth]{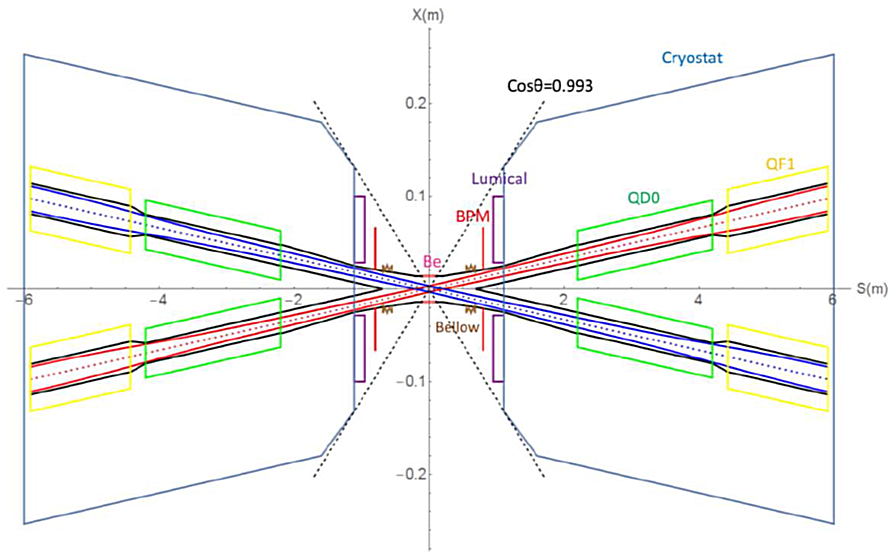}
\qquad
\includegraphics[width=.34\textwidth]{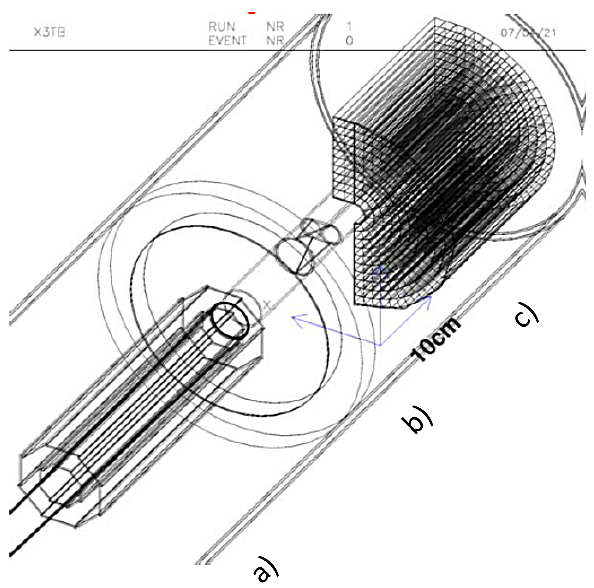}
\caption{\label{fig:1} Layout of the MDI region at CEPC (left) and a possible positioning of octagon silicon layer surrounding the beam pipe (a), silicon tracking disc (b) in front of the LYSO luminometer (c) (right).}
\end{figure}

The Machine Detector Interface (MDI) of CEPC covers the area of 6 m from the interaction point (IP) along the z-axis, in both directions. The accelerator components inside the detector without shielding are within a conical space with an opening angle of 118 mrad figure \ref{fig:1} left \cite{Suen}). The two beams collide at the IP with a crossing angle of 33 mrad in the horizontal plane with the final focus length of 2.2 m. Luminometer at CEPC is proposed to cover the polar angle region between 30 mrad and 105 mrad (with the fiducial volume between 53 mrad and 79 mrad) corresponding to the luminometer aperture of 28.5 mm for the inner radius and 100 mm for the outer, at 950 mm distance from the interaction point. Since the luminometer will be placed a half a way to the tracking volume longitudinally, shower leakage from the outer edge of the luminometer has been studied and proven to be negligible after absorption by a 5 mm iron filter positioned around the luminometer. Layout of the interaction region at CEPC, a possible positioning of and a design of the luminometer (Lu-based scintillating crystals - LYSO) are shown in figure \ref{fig:1}. Alternative design would include Si-W sandwich calorimeter with finely radially segmented Si sensors (i.e. 1.8 mm pitch). In both cases, a silicon disc (figure \ref{fig:1} right b) \cite{Suen}) will be positioned in front of the luminometer, to ensure electron-photon separation and alignment of the device. The evaluation of the impact that the energy and polar angle reconstruction have on the integrated luminosity measurement should be additionally discussed, after the final decision on luminometer design and technology is settled.

\section{Integrated luminosity measurement and systematic uncertainties}
\label{sec:sec3}

Integrated luminosity measurement is a counting experiment based on Bhabha scattering. It is defined as:  $\mathcal L=N_{Bh}/\sigma_{Bh}$, where $N_{Bh}$ is Bhabha count in the certain phase space and within the detector acceptance (fiducial) region in the certain time interval and $\sigma_{Bh}$ is the theoretical cross-section in the same geometrical and phase space. However, in a real experiment there are several effects influencing Bhabha count. Here we list the major ones, modeled in the simulation, which assumes the CEPC beams as in \cite{CEPC_CDR} and detector geometry as described in Section \ref{sec:sec2}, with the luminometer positioned at the outgoing beams. Discussed systematic effects include: uncertainties originating from mechanics (detector manufacturing, positioning and alignment); uncertainties originating from the beam properties (center-of-mass energy, beam-energy asymmetry, beam synchronization, IP displacements due to finite bunch sizes) and Bhabha miscounts from two-photon processes as a possible source of physics background. To control integrated luminosity at the required level of $10^{-4}$ ($10^{-3}$) at the $Z^{0}$ pole (240 GeV), both Bhabha count and theoretical cross-section should be known with the same precision. Further we discuss feasibility and requirements for such a precision of the Bhabha count, considering each systematic effect individually.

\subsection{Uncertainties from mechanics and positioning}
\label{sec:sec3.1}

Systematic uncertainties from detector and machine-detector interface related effects have been quantified through a simulation study, assuming $10^{7}$ Bhabha scattering events generated using BHLUMI V4.04 Bhabha  event  generator \cite{BHLUMI}, at  two CEPC center-of-mass energies: 240 GeV and $Z^{0}$ production threshold. Final state particles are generated in the polar angle range from 45 mrad to 85 mrad that is within a few mrad margin outside of the detector fiducial volume to allow events with non-collinear final state radiation to contribute. The effective Bhabha cross-section in this angular range is ~5 nb at 240 GeV and ~50 nb at the $Z^{0}$ pole. We assume that the shower leakage from the luminometer is negligible. Furthermore, we have assumed event selection in polar angle acceptance to be asymmetric between the left and right arms of the detector, as it has been done at OPAL (\cite{OPAL}, Chapter 1.3). That is, at one side we consider the full fiducial volume, while at the other side we shrink the inner radial acceptance by 1 mm. This has been done subsequently to the left (L) and right (R) sides of the luminometer, on event by event basis, resulting in cancelation of systematic uncertainties caused by the assumption of L-R symmetry of a Bhabha event.

Considered detector-related uncertainties arising from manufacturing, positioning and alignment are:
\begin{itemize}
	\item[$\bullet$]maximal uncertainty of the luminometer inner radius ($\Delta r_{in}$),
	\item[$\bullet$]RMS of the Gaussian spread of the measured radial shower position with respect to the true impact position in the luminometer front plane ($\sigma_{r}$),
	\item[$\bullet$]maximal absolute uncertainty of the longitudinal distance between left and right halves of the luminometer ($\Delta l$),
	\item[$\bullet$]RMS of the Gaussian distribution of mechanical fluctuations of the luminometer position with respect to the IP, caused by vibrations and thermal stress, radial and axial  ($\sigma_{x_{IP}}, \sigma_{z_{IP}}$),
	\item[$\bullet$]maximal absolute angular twist of the calorimeters corresponding to different rotations of the left and right detector axis with respect to the outgoing beam ($\Delta \varphi$).
\end{itemize}

\begin{figure}[htbp]
\centering 
\includegraphics[width=.47\textwidth]{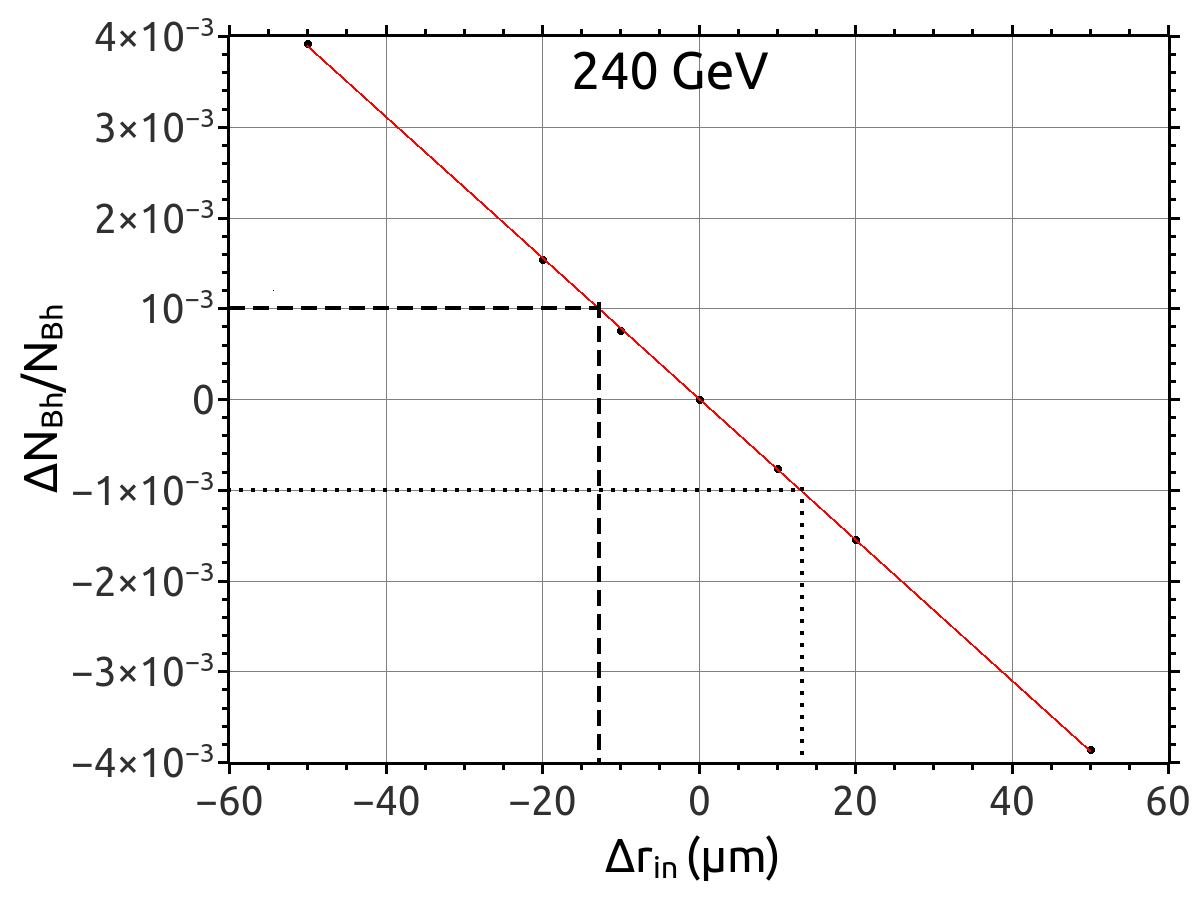}
\qquad
\includegraphics[width=.47
\textwidth]{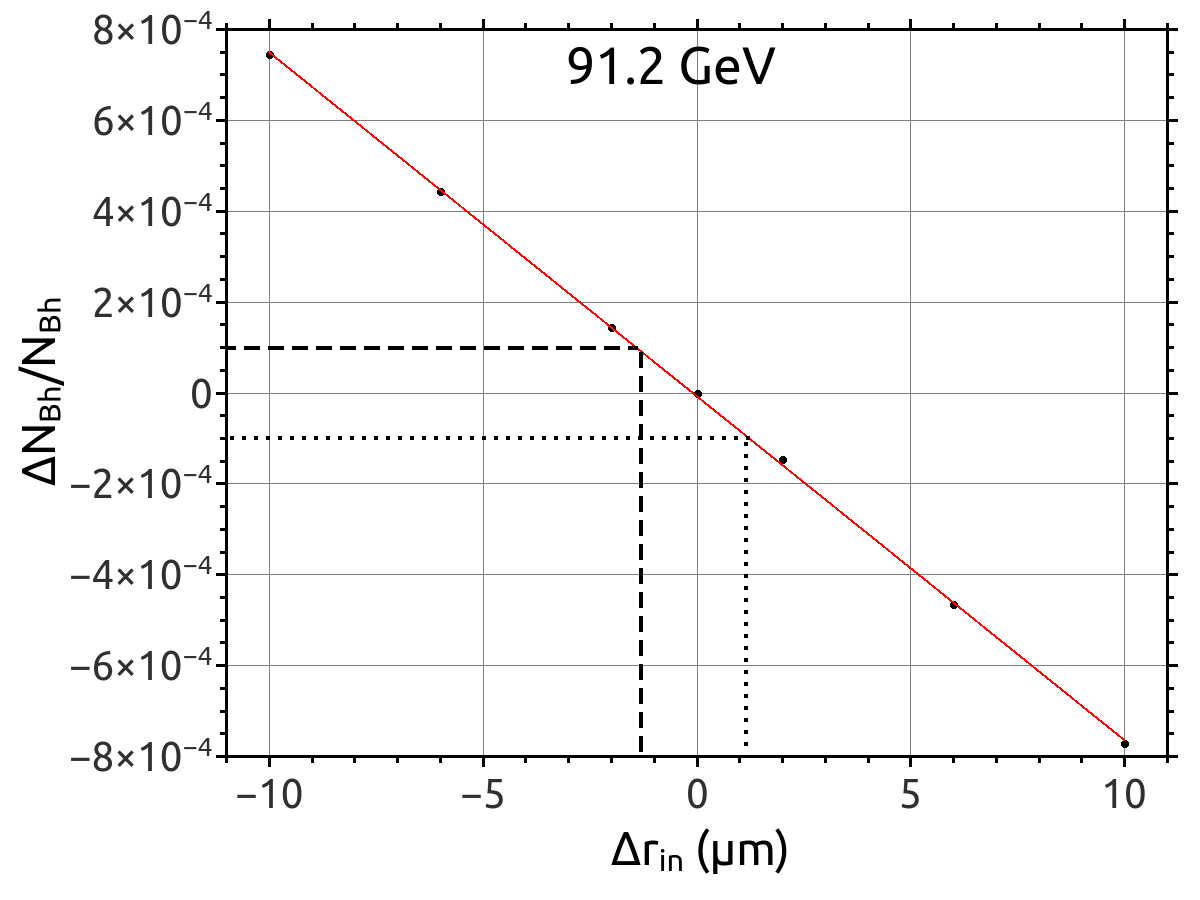}
\caption{\label{fig:2} Integrated luminosity relative uncertainty dependence on precision of the luminometer inner aperture $\Delta r_{in}$, where $\Delta r_{in} \sim \mathcal O (10$ $\mathrm{\mu m)}$ corresponds to $10^{-3}$ relative uncertainty of Bhabha count at 240 GeV (left); At 91.2 GeV (right), $\Delta r_{in} \sim \mathcal O (1$ $\mathrm{\mu m)}$ corresponds to $10^{-4}$ relative uncertainty of Bhabha count.}
\end{figure}

Considered deviations are maximal, as we assumed $10^{-3}$ and $10^{-4}$ contribution to the relative uncertainty of integrated luminosity from each individual effect, at 240 GeV and $Z^{0}$ pole respectively. Table \ref{tab:1} gives corresponding requirements of the listed parameters. Due to the $\sim 1/\theta^{3}$ dependence of the Bhabha cross-section on the polar angle, the inner aperture of the luminometer is one of the most demanding mechanical parameters to control for run at the $Z^{0}$ pole. Figure \ref{fig:2} illustrates dependence of the relative statistical precision of the integrated luminosity on the uncertainty of the inner aperture $\Delta r_{in}$\footnote{Error bars on y-axis are not given in figure \ref{fig:2}, since they are too large. To improve this, the simulated Bhabha sample should contain at least 100 times more events, which is beyond our current processing capacities}. One can notice that $\Delta r_{in}$ becomes even more sensitive if the luminometer acceptance was brought down to smaller polar angles by detector displacement or redesign.

\begin{table}[tbp]
	\centering
	\caption{\label{tab:1} Required absolute precision of mechanical parameters individually contributing to the relative uncertainty of the integrated luminosity as $10^{-3}$ ($10^{-4}$) at 240 GeV CM energy ($Z^{0}$ pole). $\Delta r_{in}$ precision is rounded to one significant figure w.r.t. the values from figure \ref{fig:2}, having in mind the statistical uncertainty of $\Delta N_{Bh}/N_{Bh}$.\\}
	\small
\begin{tabular}{|l|c|c|}
	\hline 
	\textbf{parameter} & \textbf{precision @ 240 GeV } & \textbf{precision @ 91 GeV } \\ 
	\hline 

	$\Delta r_{in}$ ($\mu$m) & 10 & 1 \\ 
	\hline 
	$\sigma_{r}$ (mm) & 1.00 & 0.20 \\ 
	\hline 
	$\Delta l$ (mm) & 1.00 & 0.08 \\ 
	\hline 
	$\sigma_{x_{IP}}$ (mm) & 1.0 & 0.5 \\ 
	\hline 
	$\sigma_{z_{IP}}$ (mm) & 10 & 7 \\ 
	\hline 
	$\Delta \varphi$ (mrad) & 6.0 & 0.8 \\ 
	\hline 
\end{tabular} 
\end{table}

\subsection{MDI related uncertainties}
\label{sec:sec3.2}

Several uncertainties that may arise from the beam properties and its delivery to the interaction point are considered:
\begin{itemize}
    \item[$\bullet$]maximal deviation ($\Delta E$) of the individual beam energy from its nominal value, resulting in asymmetry in energy of the incoming $e^{+}$ and $e^{-}$ beams (that may be caused by various effects, from the beam energy spread to beamstrahlung and initial state radiation),
	\item[$\bullet$]maximal uncertainty of the average net CM energy ($\Delta E_{CM}$) from the Bhabha cross-section calculation based on $\sigma_{Bh} \sim 1/E^{2}_{CM}$ dependence,
	\item[$\bullet$]maximal radial  ($\Delta x_{IP}^{BS}$) and axial  ($\Delta z_{IP}^{SY}$) IP position displacements with respect to the luminometer, caused by the finite transverse beam sizes and beam synchronization respectively,
	\item[$\bullet$]maximal time shift in beam synchronization ($\Delta \tau$) leading to the IP longitudinal displacement  $\Delta z_{IP}^{SY}$. 
\end{itemize}

\begin{figure}[htbp]
\centering 
\includegraphics[width=.47\textwidth]{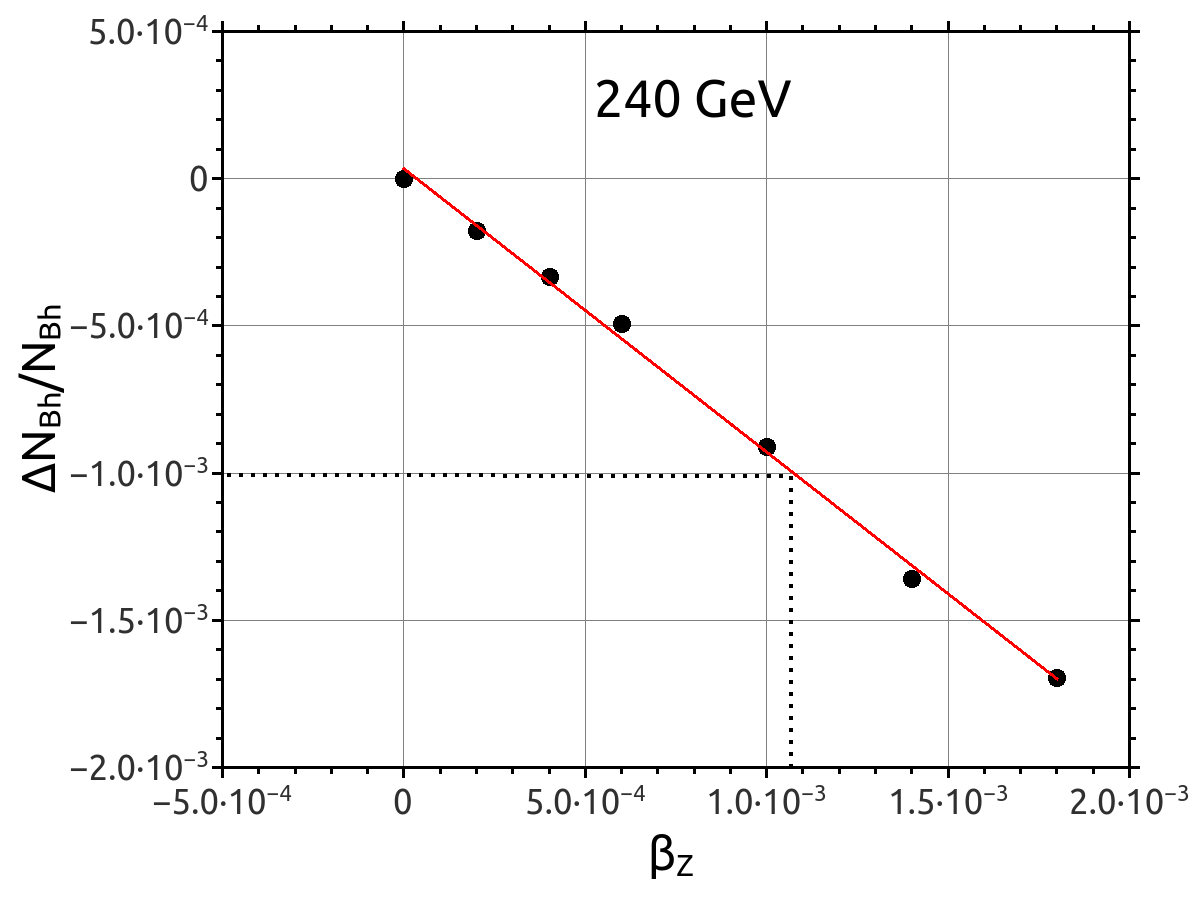}
\qquad
\includegraphics[width=.47\textwidth]{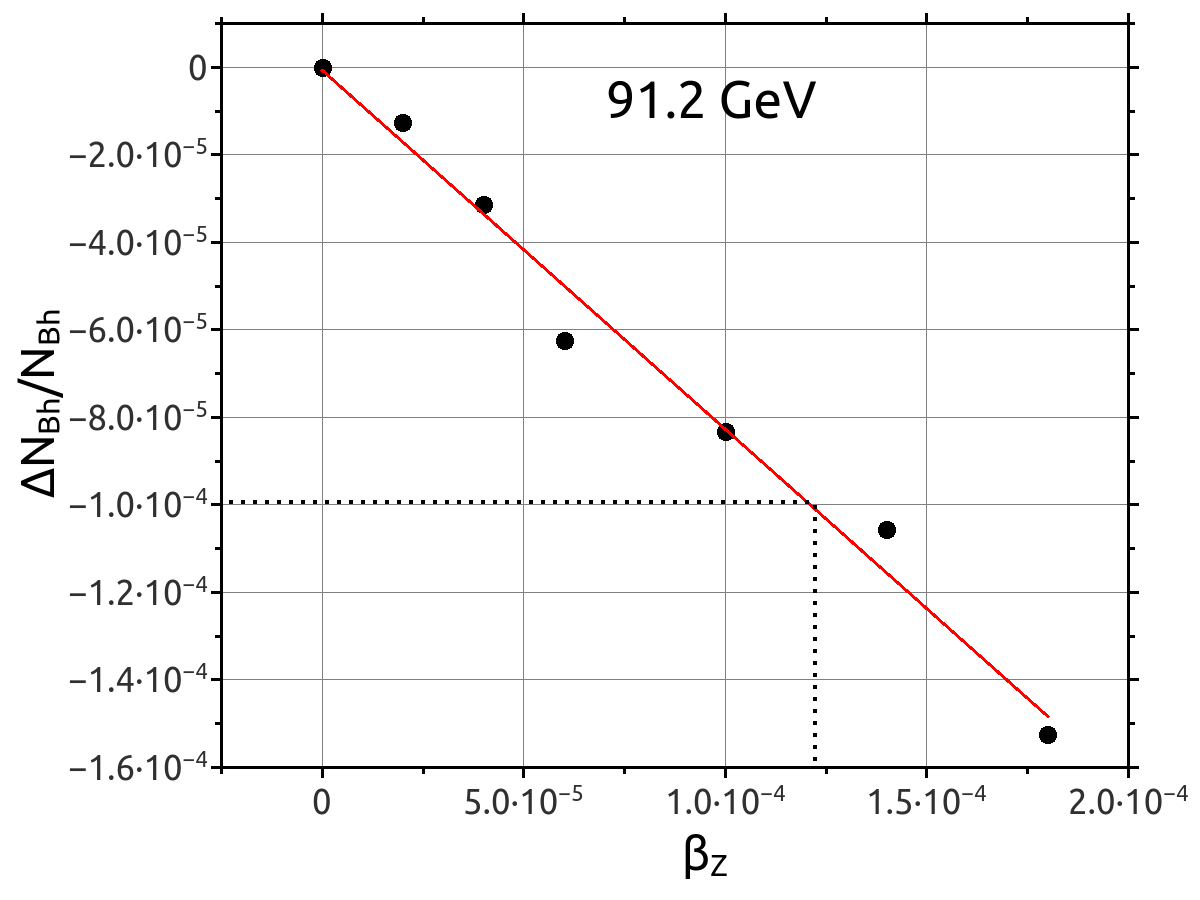}
\caption{\label{fig:3} Loss of the Bhabha count in the luminometer due to the longitudinal boost of the CM frame $\beta_{z}$, where $\beta_{z}=2 \cdot \Delta E/E_{CM}$. Event selection asymmetric in polar angle, as described in Section \ref{sec:sec3.1}, is applied. Dotted line indicates $10^{-3}$ and $10^{-4}$ relative uncertainty of the Bhabha count required at 240 GeV and $Z^{0}$ pole CEPC run, shown on the left and right respectively}
\end{figure}

Table \ref{tab:2} gives absolute uncertainties of these parameters contributing to the relative uncertainty of integrated luminosity as $10^{-3}$ ($10^{-4}$) at 240 GeV CM energy ($Z^{0}$ pole). Figure \ref{fig:3} illustrates the counting loss in luminometer due to longitudinal boost of the CM frame ($\beta_{z}=2 \cdot \Delta E/E_{CM}$) at both center-of-mass energies. $\Delta E$ values in Table \ref{tab:2} are derived from figure \ref{fig:3} and rounded with respect to statistical sizes of the samples to one and two significant figures at the $Z^{0}$ pole and 240 GeV CM energy, respectively.

The relevant challenge at the $Z^{0}$ pole comes from the fact that the uncertainty of energy of individual beams needs to be controlled at the level of $\sim 10^{-4}$ with respect to the nominal beam energy, which is apparently smaller than the foreseen BES at CEPC of 0.08\% corresponding to ~36.5 MeV. The current value of BES at the $Z^{0}$ pole will contribute to the relative uncertainty of the Bhabha count as $\sim 8 \cdot 10^{-4}$, through the asymmetry in beam energies equivalent to the longitudinal boost of the CM system of initial (final) states with respect to the laboratory frame and the consequent loss of the Bhabha coincidence due to accolinearity.

At 240 GeV CM energy conditions for precision integrated luminosity determination are more relaxed and the current BES of 0.134\% corresponding to the individual beam energy uncertainty of ~161 MeV contributes approximately as $1.3 \cdot 10^{-3}$ to the integrated luminosity uncertainty from the beam energy asymmetry. The impact of the BES uncertainty on the integrated luminosity precision will be separately discussed in Section \ref{sec:sec4.2}, once the precision of BES measurement is estimated (Section \ref{sec:sec4.1}).

\begin{table}[tbp]
	\centering
	\caption{\label{tab:2} Required absolute precision of MDI parameters contributing to the relative uncertainty of the integrated luminosity of $10^{-3}$ ($10^{-4}$) at 240 GeV CM energy ($Z^{0}$ pole). The average net center-of-mass energy uncertainty $\Delta E_{CM}$ limits are derived by error propagation from the Bhabha cross-section calculation.\\}
	\small
	\begin{tabular}{|l|c|c|}
		\hline 
		\textbf{parameter} & \textbf{precision @ 240 GeV } & \textbf{precision @ 91 GeV } \\ 
		\hline 
		$\Delta E_{CM}$ (MeV) & 120 & 5 \\ 
		\hline 
		$\Delta E$ (MeV) & 130 & 5 \\ 
		\hline 
		$\Delta x_{IP}^{BS}$ (mm) & 1.0 & 0.5 \\ 
		\hline 
		$\Delta z_{IP}^{SY}$ (mm) & 10 & 2 \\ 
		\hline 
		$\Delta \tau$ (ps) & 15 & 3 \\ 
		\hline 
\end{tabular} 
\end{table}

\subsection{Two-photon processes as a background}
\label{sec:sec3.3}

\begin{figure}[htbp]
\centering 
\includegraphics[width=.30\textwidth]{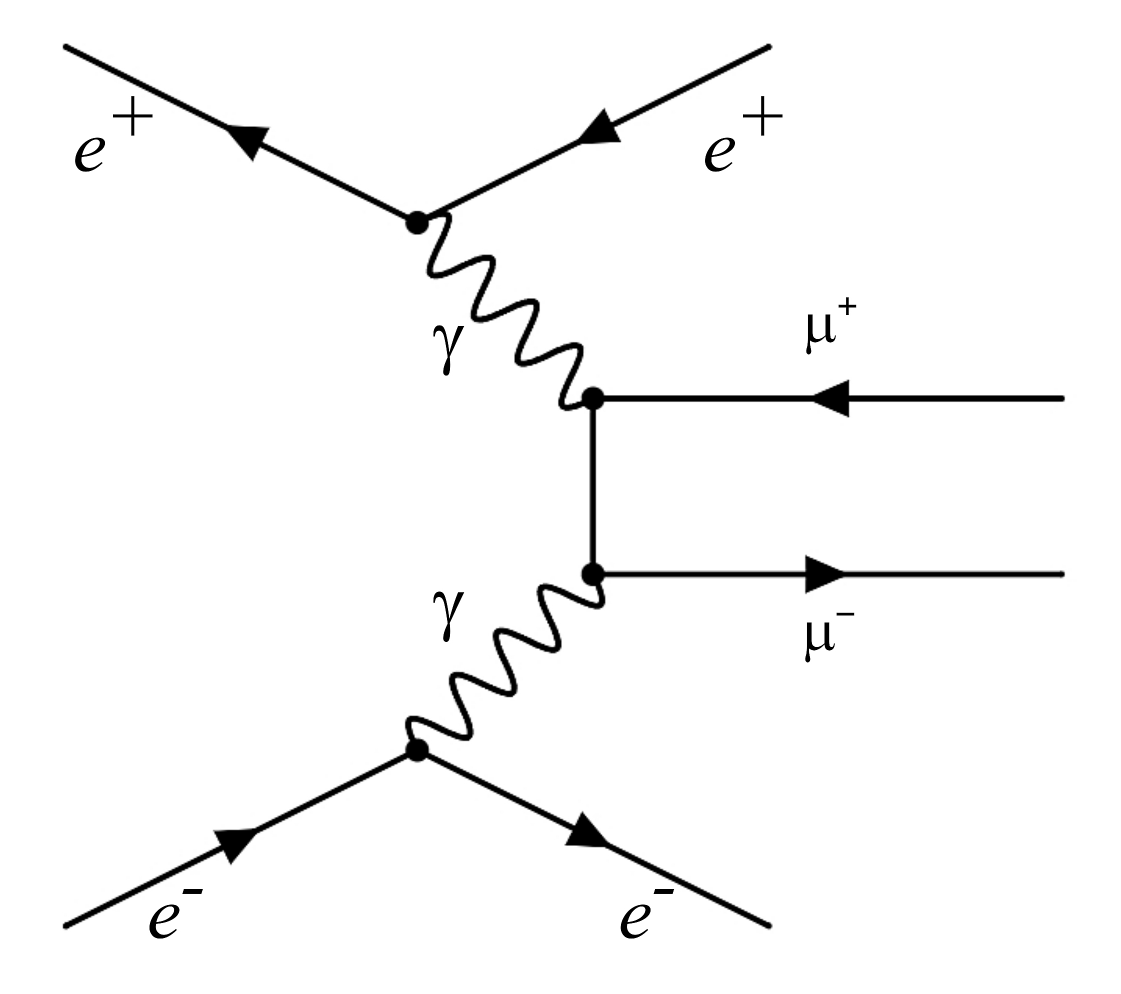}
\caption{\label{fig:4} Feynman diagram of two-photon process producing $e^{+}e^{-} \mu ^{+} \mu ^{-}$ final state in $e^{+}e^{-}$ collisions.}
\end{figure}

In $e^{+}e^{-}$ collisions there are several Feynman diagrams of four-fermion final state processes (multiperipheral, annihilation, brehmstrahlung and conversion) contributing to possible $e^{+}e^{-}f\bar{f}$ final state background for the Bhabha scattering. The multiperipheral (two-photon) processes (given in figure \ref{fig:4}) are considered, due to the large cross-section ($\sim$nb) and the fact that spectator electrons (positrons) are emitted at very small polar angles. Even though the most of high-energy spectators will go below the luminometer’s angular acceptance region, some of them can still be misidentified as Bhabha electrons, in particular at  240 GeV, where the ratio of signal and background cross-sections as a function of $s$ ($s=E_{CM}^{2}$) disfavors Bhabha scattering at higher center-of-mass energies: $\sigma_{Bh}/\sigma_{2\gamma} \sim 1/(s \cdot ln(s))$. Here, we quantify the contribution from this source of physics background, assuming geometrical parameters as in Section \ref{sec:sec2}.

\begin{figure}[htbp]
\centering 
\includegraphics[width=.47\textwidth]{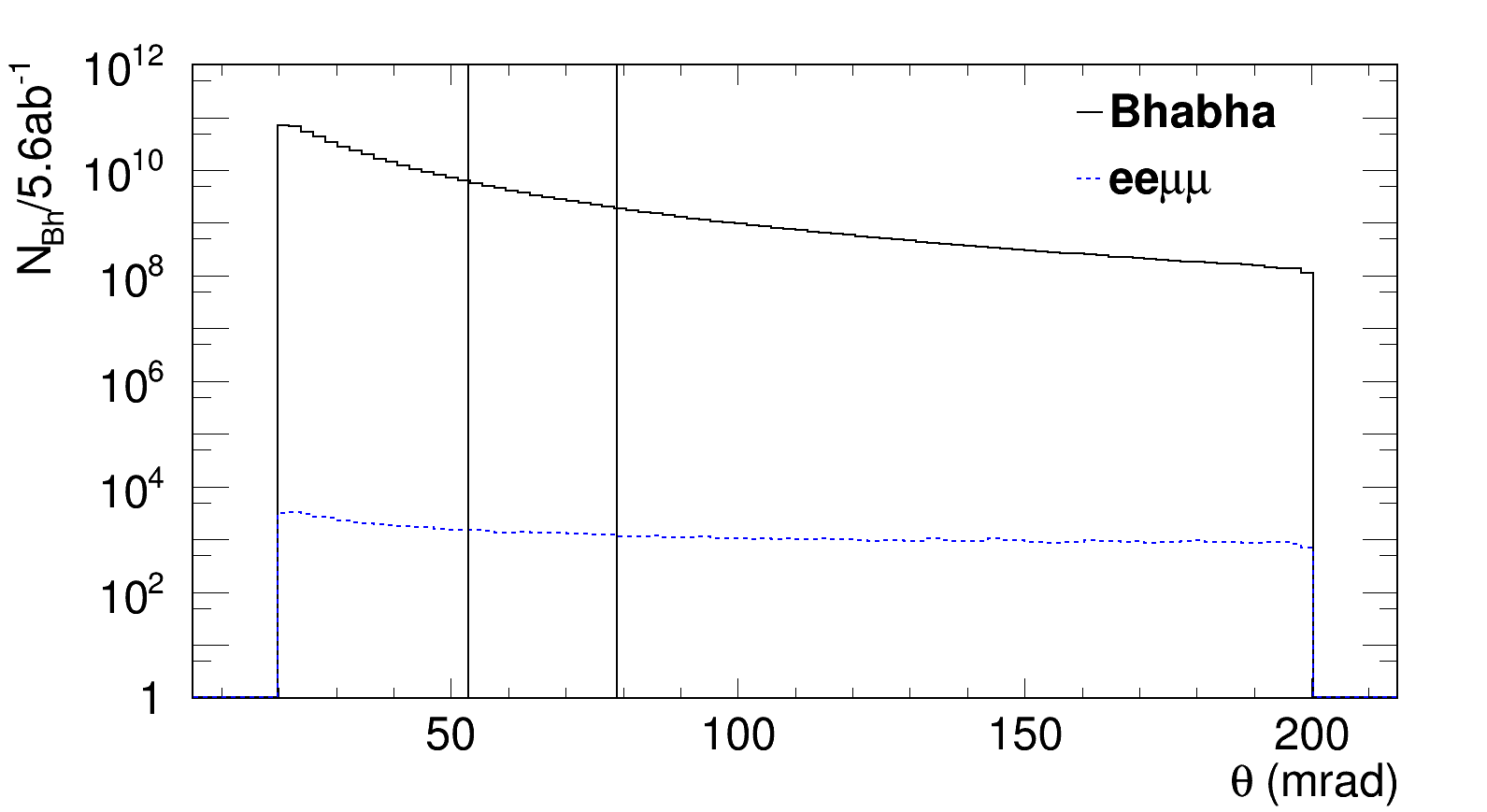}
\caption{\label{fig:5} Normalized polar angle distribution of signal (full line) and two-photon background (dotted line) events at 240 GeV, with two vertical lines indicating the luminometer’s fiducial volume.}
\end{figure}

To estimate the background to signal ratio at 240 GeV CEPC, we simulated $10^{5}$ $e^{+}e^{-} \rightarrow e^{+}e^{-} \mu ^{+} \mu ^{-}$ events using WHIZARD V2.8 \cite{WHIZARD}, with the effective cross-section $\sigma_{\mathrm{eff}}\sim 0.3$ pb in the fiducial volume of the luminometer. To illustrate cross-section dependence on the polar angle, $10^{7}$ Bhabha events are simulated using BHLUMI V4.04 in the polar angle range 20 mrad < $\theta$ < 200 mrad, with the effective cross-section of $\sim 3.3$ nb in the fiducial volume of the luminometer (figure \ref{fig:5}). Counts are normalized to 5.6 $\mathrm{ab^{-1}}$ of integrated luminosity, corresponding to 7 years of data taking at 240 GeV CEPC. 

The figure \ref{fig:5} illustrates that most of the two-photon spectators go below the luminometer acceptance, while the contamination of the signal is significantly below $10^{-4}$ in the luminometer’s fiducial volume, even without any event selection. The total amount of background should be conservatively scaled by a factor 3 to account for lepton flavor integration.

In general, background suppression is possible with the coplanarity requirement of Bhabha electrons (positrons) identified in left and right detector arms, based on azimuthal angle ($\varphi$) measurements ($|\varphi_{e^{+}}-\varphi_{e^{-}}|$). Event selection based on the relative energy $E_{rel}=\frac{|E_{e^{+}}+E_{e^{-}}|}{2\cdot E_{beam}}$ is additionally useful to suppress off-momentum particles. Finally, physics background can be always taken as a correction to the Bhabha count whenever the uncertainty of its cross-section is available.

\section{Impact of the beam energy spread}
\label{sec:sec4}

Uncertainty of the beam energy spread determination will contribute to the overall systematic uncertainty of the integrated luminosity measurement affecting the asymmetry of beam energies and consequently providing longitudinal boost $\beta_{z}$ of the colliding system in the laboratory frame. Below we discuss feasibility of precision BES measurement at CEPC (Section \ref{sec:sec4.1}) and its impact on the integrated luminosity precision (Section \ref{sec:sec4.2}). In addition, having in mind that precision electroweak measurements at the $Z^{0}$ pole are among priorities of future electron-positron colliders’ physics program, we discuss the impact of estimated BES precision on electroweak measurements at the $Z^{0}$ pole: $Z^{0}$ production cross-section, mass and width (Section \ref{sec:sec4.2}).

\subsection{Method of the beam energy spread determination}
\label{sec:sec4.1}

Motivated by the similar work done at FCCee \cite{Janot} and having in mind that numerous precision observables, including integrated luminosity, depend on the precision of BES, we looked into possibility to measure it at CEPC using well defined central process, such as di-muon production $e^{+}e^{-} \rightarrow \mu ^{+} \mu ^{-}$. Having in mind projected performance of the CEPC's central tracker to reconstruct muons efficiently (~99\%) and precisely ($\Delta p_{t}/p_{t}^{2} \sim 10^{-5}$ $\mathrm{GeV^{-1}}$), the process $e^{+}e^{-} \rightarrow \mu ^{+} \mu ^{-}$ with a cross-section of 1.5 nb at the $Z^{0}$ pole seems to be an optimal choice.

\begin{figure}[htbp]
\centering 
\includegraphics[width=.47\textwidth]{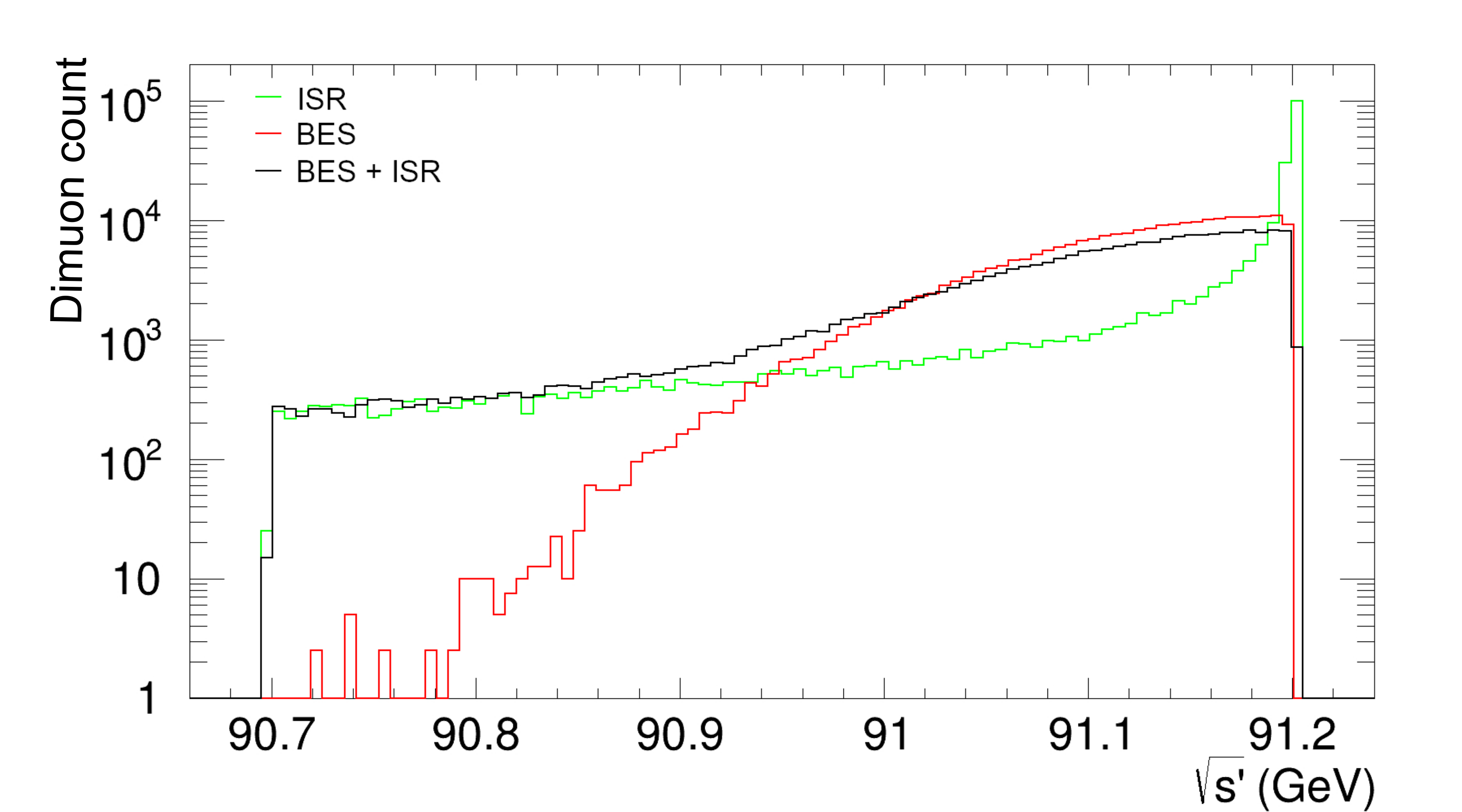}
\qquad
\includegraphics[width=.47\textwidth]{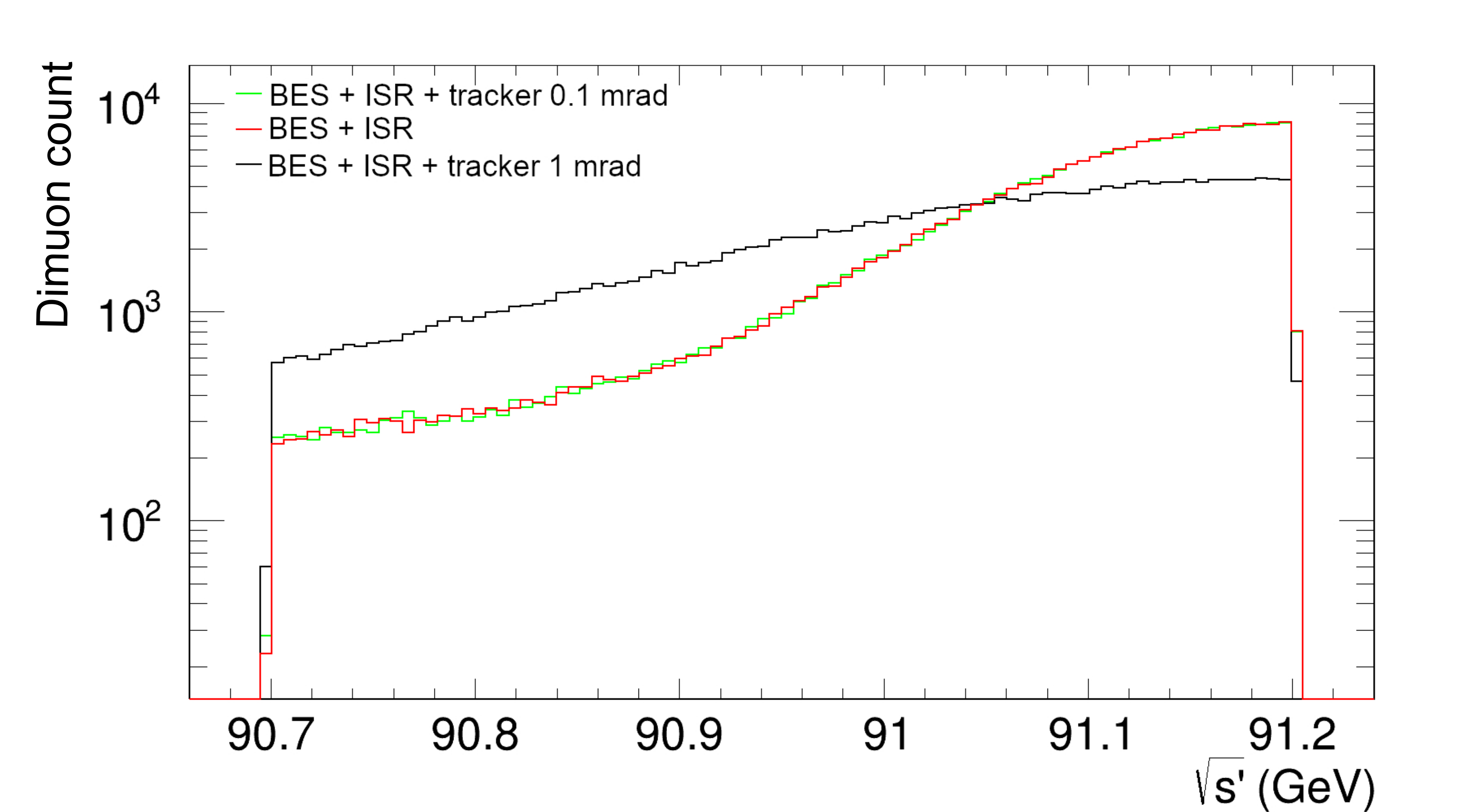}
\caption{\label{fig:6} Count of di-muon events versus the effective CM energy (top part of the spectrum) at the $Z^{0}$ pole. BES is the dominant effect to reduce the number of events at the maximal CM energy (left); $\sqrt{s'}$ sensitivity to the BES with finite central tracker resolution (right).}
\end{figure}

We argue that the effective CM energy ($\sqrt{s'}$) is sensitive to variation of the BES that consequently can be determined from the population of the peak of the $\sqrt{s'}$ distribution. To determine $\sqrt{s'}$ sensitivity to the BES, we generated several hundred thousand $e^{+}e^{-} \rightarrow \mu ^{+} \mu ^{-}$ events at 91.2 GeV and 240 GeV CM energies. Events are generated using WHIZARD V2.8, in the polar angle range from $8^{o}$ to $172^{o}$, which corresponds to the angular acceptance of the central tracker (TPC) at CEPC. In the simulated events, the effects like the initial state radiation (ISR) and detector angular resolution are modeled and studied individually, to evaluate their impact on the $\sqrt{s'}$ distribution with respect to the concurrent BES. Detector energy resolution is simulated by performing Gaussian smearing of the muons’ polar angles. Applying 0.1 mrad smearing corresponds to 100 $\mathrm{\mu m}$ position resolution foreseen for TPC at CEPC. 
The effective CM energy squared, s’, can be calculated from the reconstructed muons’ polar angles \cite{OPAL2}, as:
\begin{equation}
\label{eq:1}
\frac{s'}{s}=\frac{\sin{\theta^{+}}+\sin{\theta^{-}}-|\sin{\theta^{+}+\theta^{-}}|}{\sin{\theta^{+}}+\sin{\theta^{-}}+|\sin{\theta^{+}+\theta^{-}}|},
\end{equation}
relying on the excellent TPC spatial resolution. As illustrated in figure \ref{fig:6} (left), BES dominates the $\sqrt{s'}$ shape at energies close to the nominal CM energy, while the figure \ref{fig:6} (right) illustrates the effect of muon polar angle resolutions of 0.1 mrad and 1 mrad on top of ISR and the BES. From figure \ref{fig:6} it is clear that 0.1 mrad central tracker resolution does not affect the $\sqrt{s'}$ sensitivity to the BES. On the other hand, tracker resolution of 1 mrad significantly influences the method. We found that polar angle resolution in central tracker should not be larger than 0.5 mrad, corresponding to the 500 $\mathrm{\mu m}$ position resolution. The same holds for 240 GeV CEPC run.

\begin{figure}[htbp]
\centering 
\includegraphics[width=.47\textwidth]{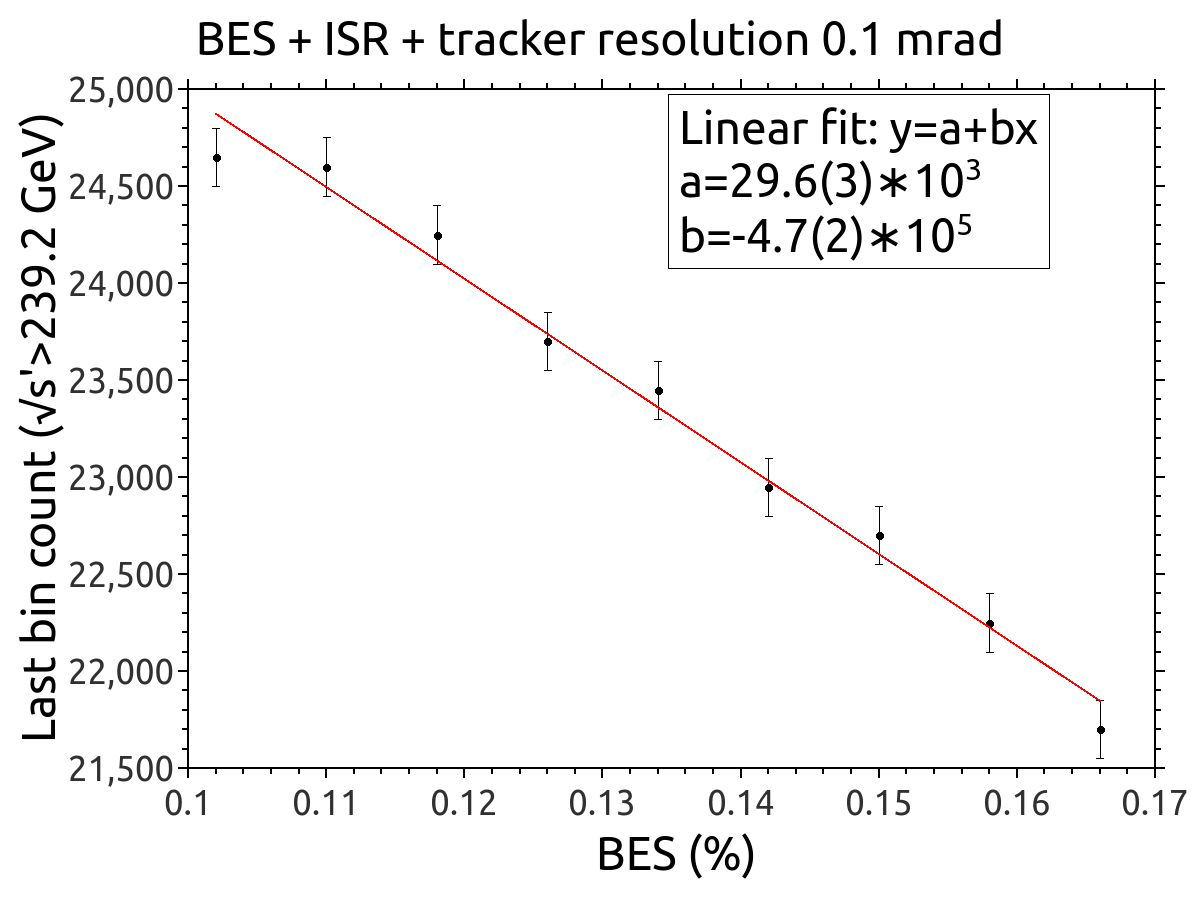}
\qquad
\includegraphics[width=.47\textwidth]{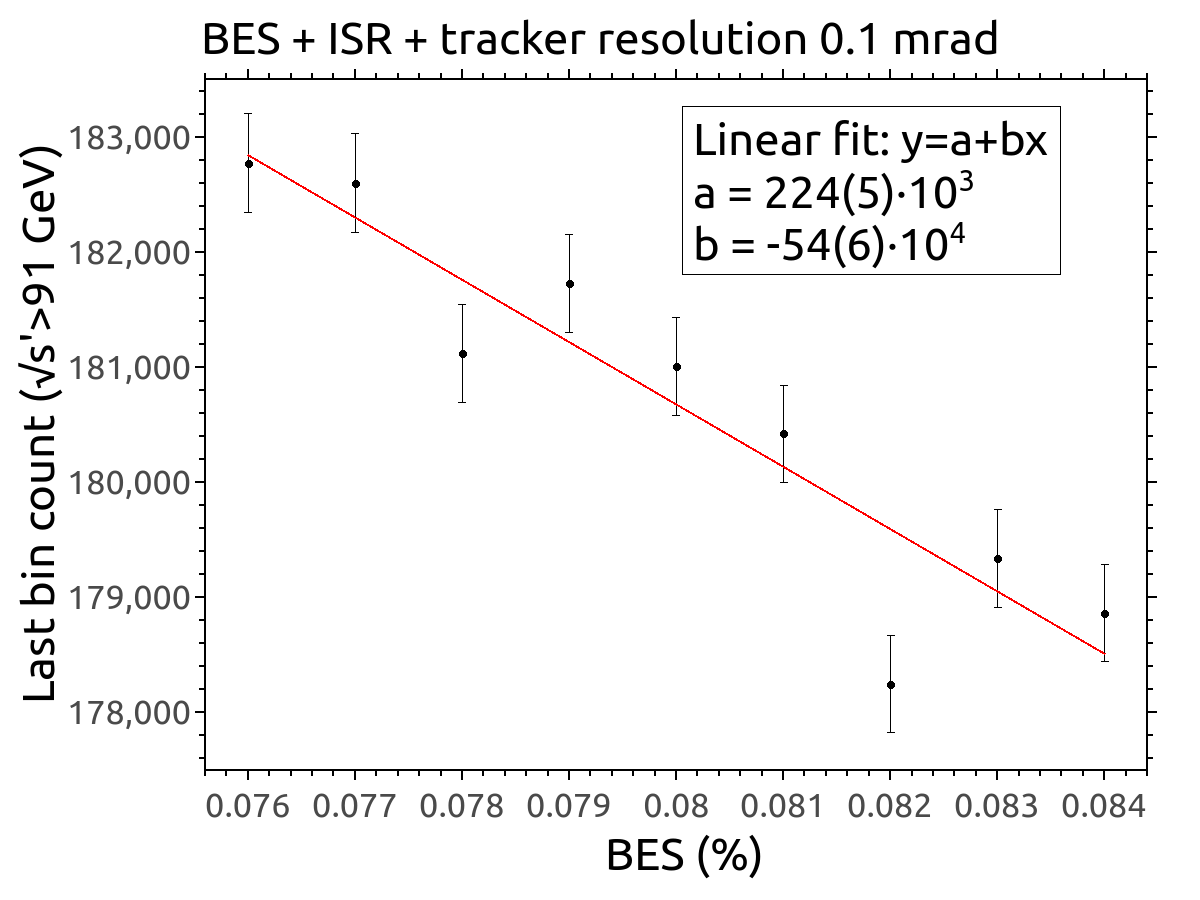}
\caption{\label{fig:7} Number of di-muon events in the top 3‰ of the nominal CM energy at 240 GeV (left) and in the top 2‰ of the nominal CM energy at 91.2 GeV (right) for various BES values.}
\end{figure}

To exploit $\sqrt{s'}$ peak sensitivity to the BES values, BES is varied around the nominal value, generating $10^{5}$ ($2.5 \cdot 10^{5}$) events per BES variation at 240 GeV (91.2 GeV). The observed dependence is illustrated in figure \ref{fig:7}, at the 240 GeV (left) and $Z^{0}$ pole (right). As expected, larger BES leads to the larger reduction of the number of di-muon events carrying near to maximal available energy from the collision. Knowing this dependence from simulation allows for determination of the effective BES (denoted as $\delta '$) once the count of di-muon events is known experimentally.

Table \ref{tab:3} shows that 1.2\% relative statistical uncertainty of the BES measurement arises after only 3 minutes of data taking with $1.02\cdot10^{36}$ $\mathrm{cm^{-2} s^{-1}}$ instantaneous luminosity at the $Z^{0}$ pole. BES can be measured with the total relative uncertainty of 25\%, where the systematic contribution comes from the calibration curve (figure \ref{fig:7}, right). The total uncertainty is obtained  by combining statistical and systematic components as uncorrelated. At 240 GeV center-of-mass energy, BES can be measured with 15\% total uncertainty (figure \ref{fig:7}, left) and 2.3\% relative statistical uncertainty in approximately 5 days of data taking with instantaneous luminosity of $5.2 \cdot 10^{34}$ $\mathrm{cm^{-2} s^{-1}}$. Total uncertainty of the BES translates into maximal uncertainty of individual beam energies $\Delta E_{BES}$ of 9 MeV (24 MeV) at the $Z^{0}$ pole (240 GeV).

\begin{table}[!htp]
	\centering
	\caption{\label{tab:3}BES relative variations experimentally accessible at CEPC. Values that are calculated or obtained from simulated BES measurement are bolded. Other entries in the table are taken from \cite{CEPC_CDR}. Total relative uncertainty of BES includes statistical and systematic uncertainties summed as uncorrelated. Total relative uncertainty of BES determination contributes to the absolute uncertainty of the beam energy as $\Delta E_{BES}$.\\}
	\footnotesize
	\begin{tabular}{|c|c|c|c|c|c|c|c|c|}
		\hline 
		CEPC & \vtop{\hbox{\strut $\mathcal{L}$ @ IP} \hbox{\strut ($\mathrm{cm^{-2}s^{-1}}$)}} & \vtop{\hbox{\strut Nominal} \hbox{\strut BES}  \hbox{\strut $\delta$ (\%)}} & \vtop{\hbox{\strut Number} \hbox{\strut of} \hbox{\strut events}} & \vtop{\hbox{\strut Cross-} \hbox{\strut section} \hbox{\strut$e^{+}e^{-}$} \hbox{\strut$\rightarrow \mu^{+}\mu^{-}$}} & \vtop{\hbox{\strut Collecting} \hbox{\strut time}} &  \vtop{\hbox{\strut Relative} \hbox{\strut statistical} \hbox{\strut uncertainty} \hbox{\strut of BES}} & \vtop{\hbox{\strut Total} \hbox{\strut relative} \hbox{\strut uncertainty} \hbox{\strut of BES}} & \vtop{\hbox{\strut $\Delta E_{BES}$} \hbox{\strut (MeV)}} \\ 
		\hline 
		$Z^{0}$ pole & $1.02 \cdot 10^{36}$ & 0.080 & $2.5 \cdot 10^{5}$ & \textbf{1.5 nb} & \textbf{3 min} & \textbf{1.2\%} & \textbf{25\%} & \textbf{9} \\ 
		\hline 
		240 GeV & $5.2 \cdot 10^{34}$ & 0.134 & $1.0 \cdot 10^{5}$ & \textbf{4.1 pb} & \textbf{5 days} & \textbf{2.3\%} & \textbf{15\%} & \textbf{24} \\		
	\hline
\end{tabular} 
\end{table}

\subsection{Impact on integrated luminosity measurement and precision electroweak  observables}
\label{sec:sec4.2}

\begin{figure}[htbp]
\centering 
\includegraphics[width=.47\textwidth]{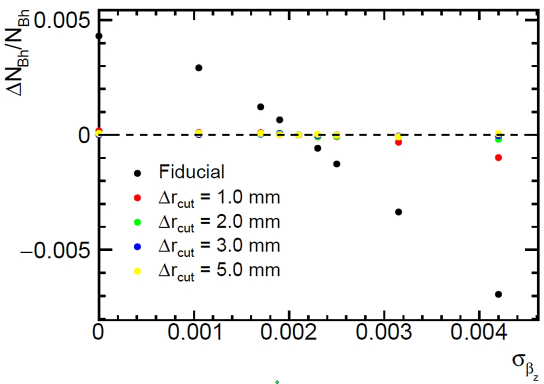}
\caption{\label{fig:8} Sensitivity of the Bhabha count on the longitudinal boost $\sigma_{\beta_{z}}$ caused by BES uncertainty, for counting in the symmetrical fiducial volume and LEP-style selection ($\Delta r \ne 0$) at the $Z^{0}$ pole.}
\end{figure}

As already mentioned is Section \ref{sec:sec3.2}, asymmetry in beam energies will give a rise to the longitudinal boost $\beta_{z}$ leading to the loss of coincidence of Bhabha hits in left and right arms of the luminometer.  Considering BES uncertainty individually as a source of longitudinal boost $\sigma_{\beta_{z}}$ ($\sigma_{\beta_{z}}=2 \cdot \Delta E_{BES}/E_{CM}$), achievable BES uncertainty of 9 MeV at the $Z^{0}$ pole will translate to $\sigma_{\beta_{z}} \sim 2 \cdot 10^{-4}$, contributing to the relative systematic uncertainty of the Bhabha count as $4 \cdot 10^{-3}$. The above holds for the symmetrical counting in the fiducial volume, while if asymmetric (LEP-style) selection described in Section \ref{sec:sec3.1} is applied, luminosity determination is practically insensitive to the precision of BES, as illustrated at figure \ref{fig:8}. The above suggests that the luminometer should be positioned at the outgoing beams. 

\begin{figure}[htbp]
\centering 
\includegraphics[width=.47\textwidth]{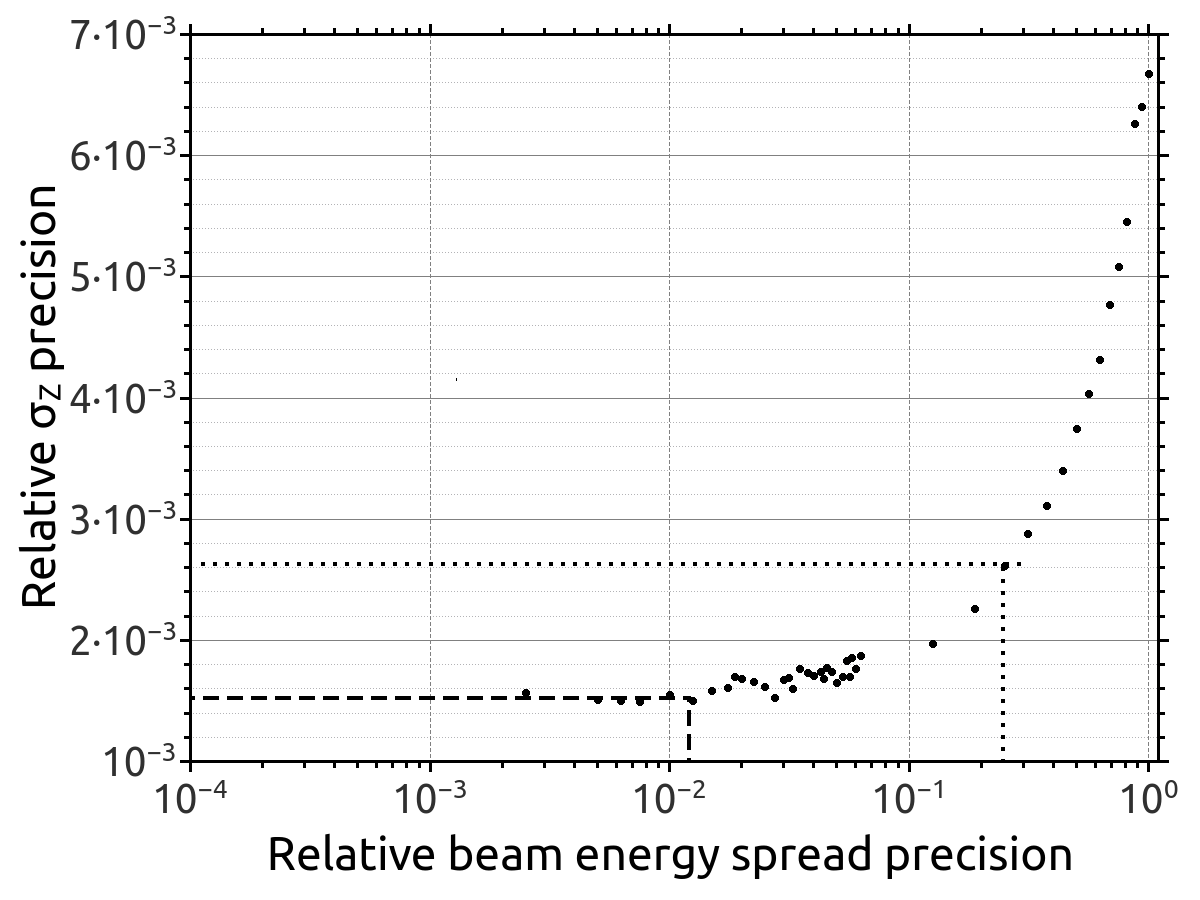}
\caption{\label{fig:9} Impact of the relative precision of the BES on the $Z^{0}$ production cross-section $\sigma_{Z}$. Impact of the BES statistical and total uncertainties are indicated with dashed and dotted lines, respectively. Error bars on ordinate are within dots.}
\end{figure}

\begin{figure}[htbp]
\centering 
\includegraphics[width=.47\textwidth]{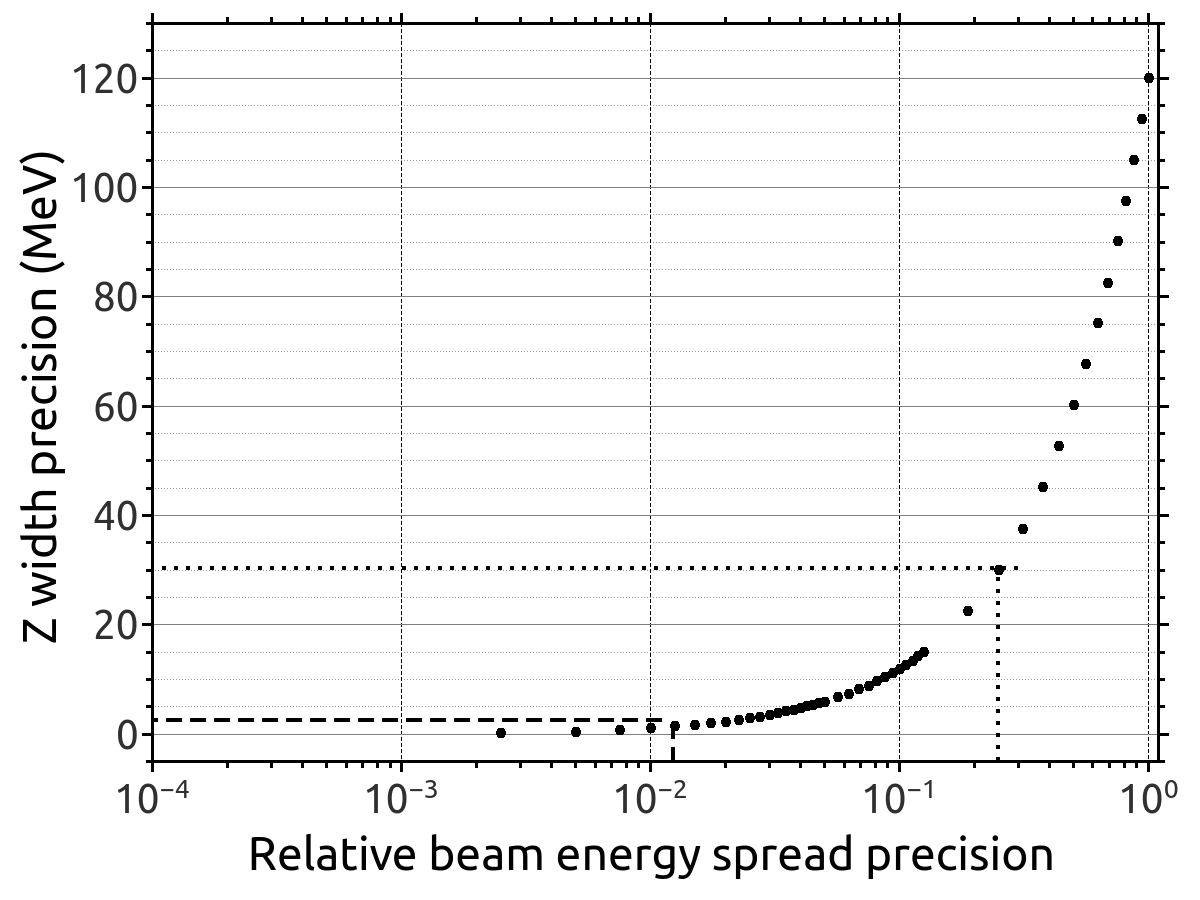}
\qquad
\includegraphics[width=.47\textwidth]{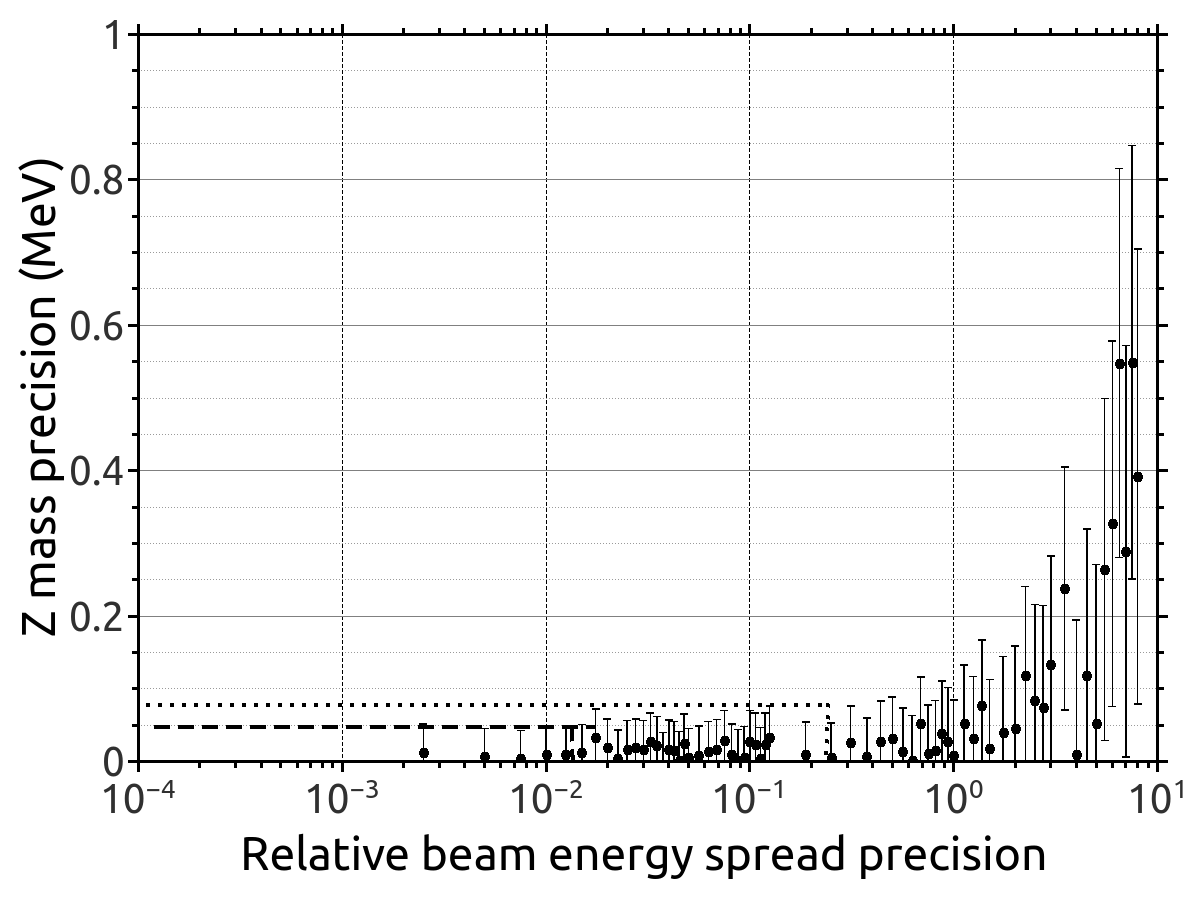}
\caption{\label{fig:10} Impact of the relative precision of the BES on the $Z^{0}$ total width (left) and mass (right) absolute precisions. Impact of the BES statistical and total uncertainties are indicated with dashed and dotted lines, respectively. In the case of $Z^{0}$ mass, precision estimate is conservatively taken to include the error bars corresponding to the standard error of the mean obtained on the sample of one million di-muon events for each beam energy spread deviation. For the $Z^{0}$ total width precision, error bars on ordinate are within dots.}
\end{figure}

As previously mentioned, several precision electroweak observables at the $Z^{0}$ pole depend on the BES uncertainty. Figures \ref{fig:9} and \ref{fig:10} illustrate that the cross-section for $Z^{0}$ production ($\sigma_{Z}$), $Z^{0}$ total width ($\Gamma_{Z}$) and mass ($m_{Z}$) will receive following contributions from the total BES uncertainty: $\delta(\sigma_{Z})\sim2.6\cdot10^{-3}$, $\Delta \Gamma_{Z}\sim30$ MeV and $\Delta m_{Z}<100$ keV, respectively. Naturally, uncertainties originated solely from the statistical uncertainty of the BES are significantly smaller, as indicated in figures \ref{fig:9} and \ref{fig:10}, corresponding to $\delta(\sigma_{Z})\sim1.5\cdot10^{-3}$, $\Delta \Gamma_{Z}\sim2$ MeV and $\Delta m_{Z}\sim50$ keV. These results are summarized in Table \ref{tab:4}, together with the BES precision impact on integrated luminosity uncertainty.

\begin{table}[tbp]
	\centering
	\caption{\label{tab:4} Impact of the BES total and statistical uncertainties on precision observables at the $Z^{0}$ pole: cross-section for the $Z^{0}$ production ($\sigma_{Z}$), $Z^{0}$ total width ($\Gamma_{Z}$) and mass ($m_{Z}$) and on integrated luminosity precision for counting in the fiducial volume and asymmetric counting.\\}
	\small
	\begin{tabular}{|l|c|c|c|c|c|}
		\hline 
		BES @ the $Z^{0}$ pole & $\delta(\sigma_{Z})$ & $\Delta\Gamma_{Z}$ (MeV) & $\Delta m_{Z}$ (keV) & $(\Delta\mathcal{L}/\mathcal{L})_{fid}$ & $(\Delta\mathcal{L}/\mathcal{L})_{asym}$\\ 
		\hline 
		Total uncertainty (25\%) & $2.6\cdot10^{-3}$ & 30 & <100 & $4\cdot10^{-3}$ & $\precsim 10^{-4}$\\ 
		\hline 
		Statistical uncertainty (1.2\%) & $1.5\cdot10^{-3}$ & 2 &50 & $2\cdot10^{-4}$ & $\precsim 10^{-4}$\\ 
		\hline
\end{tabular} 
\end{table}

\section{Conclusions}
\label{sec:sec5}

Although the method of integrated luminosity has already been studied in great detail at LEP, proposals for new $e^{+}e^{-}$ colliders call for quantification of achievable luminosity precision in each individual case. We do it here for CEPC, from the perspective of mechanical and MDI requirements, in parallel with the BES determination from the di-muon production and its impact on precision of integrated luminosity and precision electroweak observables at the $Z^{0}$ pole.

At the $Z^{0}$ pole, control of the luminometer inner radius at the micrometer level is posing the most demanding requirement regarding detector manufacturing and positioning precision. It is important to note that a modification of the luminometer inner aperture toward smaller polar angles will require control of the luminometer inner radius below micrometer precision. Another challenge comes from the uncertainty of the beam energy that might be caused by multiple factors and has to be known below the foreseen beam spread in order to contribute to the relative uncertainty of integrated luminosity as $1\cdot10^{-4}$. With the current beam design, beam energy spread that causes the beam energy asymmetry contributes with $\sim 8\cdot10^{-4}$ to relative uncertainty of the integrated luminosity at the $Z^{0}$ pole. Effective net center-of-mass energy for the cross-section calculation has to be known within a few MeV uncertainty.  Apparently, sub-per mill integrated luminosity precision at the $Z^{0}$ pole is still an open question even with the post-CDR beam properties. 

Per mill precision of the integrated luminosity measurement at 240 GeV CEPC seems to be feasible from the point of view of existing technologies and foreseen beam properties. Contribution to the integrated luminosity uncertainty from the beam energy spread does not exceed $\sim 1.3\cdot10^{-3}$.

It is also shown that with the CEPC post-CDR instantenious luminosity upgrade, beam energy spread can be determined with the total accuracy corresponding to 9 MeV beam energy uncertainty in only 3 minutes of data-taking of $e^{+}e^{-} \rightarrow \mu^{+}\mu^{-}$ events at the $Z^{0}$ pole. The accuracy is dominated by the systematic uncertainty of the method. The total precision of the BES determination has negligible effect on the integrated luminosity precision with the LEP-style counting, requiring luminometer positioned at the outgoing beams. However, BES uncertainty translates to the relative uncertainty of the $Z^{0}$ production cross-section of $2.6\cdot10^{-3}$ and absolute precisions of the $Z^{0}$ mass and width below 100 keV and 30 MeV, respectively.


\acknowledgments

This research was funded by the Ministry of Education, Science and Technological Development of the Republic of Serbia and by the Science Fund of the Republic of Serbia through the Grant No. 7699827, IDEJE HIGHTONE-P. This paper also contains results, namely figure \ref{fig:8}, accomplished in previous collaboration with S. Lukic, within the Vinca Institute HEP Group.


\end{document}